\documentclass
[preprint,prd,twocolumn,10pt,nofootinbib,noshowkeys,noshowpacs]{revtex4}%
\usepackage{amsfonts}
\usepackage{amsmath}
\usepackage{amssymb}
\usepackage{graphicx}%
\providecommand{\U}[1]{\protect\rule{.1in}{.1in}}

\begin{document}
\title{On the stabilization of the Friedmann Big Bang by the shear stresses}
\author{V. A. Belinski}
\affiliation{ICRANet, 65122 Pescara, Italy; Phys. Dept., Rome University "La Sapienza",
00185 Rome, Italy; IHES, F-91440 Bures-sur-Yvette, France.}

\begin{abstract}
The window is found in the space of the free parameters of the theory of
viscoelastic matter for which the Friedmann singularity is stable. Under
stability we mean that in the presence of the shear stresses the
\textit{generic} solution of the equations of relativistic gravity possessing
the isotropic, homogeneous and thermally equilibrated cosmological singularity exists.

\end{abstract}
\maketitle

\section{ Introduction}

Observations show that the early Universe was isotropic, homogeneous and
thermally balanced. A number of authors \cite{Bar,Pen,Goo} expressed the point
of view that also the initial cosmological singularity should be in conformity
with these properties, that is should be of the Friedmann type. But it is well
known that the Friedmann singularity for the conventional types of matter is
unstable which means that space-time cannot start isotropic expansion unless
an artificial fine tuning of unknown origin. This instability is due to the
sharp anisotropy which develops unavoidably near the generic cosmological
singularity. However, an intuitive understanding suggests that anisotropy can
be damped down by the shear viscosity which being taking into account might
results in the \textit{generic} solution with isotropic Big Bang. To search an
analytical realization of such a possibility there would be inappropriate to
use just the Eckart \cite{Eck} or Landau-Lifschitz \cite{LL} approaches to the
relativistic hydrodynamics with dissipative processes. These theories are
valid provided the characteristic times of the macroscopic motions of the
matter are much bigger than the time of relaxation of the medium to the
equilibrium state. It might happen that this is not so near the cosmological
singularity since all characteristic macroscopic times in this region tend to
zero in which case one need a theory which takes into account the Maxwell's
relaxation times on the same footing as all other transport coefficients. In a
literal sense such a theory does not exists, however, it can be constructed in
an approximate form for the cases when a medium do not deviates too much from
equilibrium and relaxation times do not exceed noticeably the characteristic
macroscopic times . It is reasonable to expect that these conditions will be
satisfied automatically for a generic solution (if it exists) near isotropic
singularity describing the beginning of the thermally balanced Friedmann
Universe accompanying by the arbitrary infinitesimally small corrections.

The main target of the efforts of many authors (starting from the first idea
of Cattaneo \cite{Cat} up to the final formulation of the generalized
relativistic theory by Israel and Stewart \cite{Isr1,Isr2}) was to bring the
theory into line with relativistic causality, that is to eliminate the
supraluminal propagation of the thermal and viscous excitations. The existence
of such supraluminal effects was the main stumbling-block for the Eckart's and
Landau-Lifschitz's descriptions of dissipative fluids. One of the first
applications of the Israel-Stewart theory to the problems of cosmological
singularity was undertaken in the article \cite{BNK}. Already in this paper
the stability of the Friedmann models under the influence of the shear
viscosity has been investigated and it was found that relativistic causality
and stability of the Friedmann singularity are in contradiction to each other.
Then the final conclusion was: "\textit{...relativistic causality precludes
the stability of isotropic collapse. The isotropic singularity cannot be the
typical initial or final state.}" However, in the present paper it will be
shown that this "no go" conclusion was too hasty since it was the result of
too restricted range for the dependence of the shear viscosity coefficient on
the energy density. As usual, in the vicinity to the singularity where the
energy density $\varepsilon$ diverges we approximate the coefficient of
viscosity $\eta$ by the power law asymptotics $\eta\sim\varepsilon^{\nu}$ with
some exponent $\nu.$ In the article \cite{BNK} (due to some more or less
plausible thoughts) we choose the values of this exponent from the region
$\nu>1/2.$ For these values of $\nu$ the negative result of paper \cite{BNK}
remains correct, but recently it made known that the boundary value $\nu=1/2$
leads to the dramatic change of the state of affairs. It turns out that for
this case there exists a window in the space of the free parameters of the
theory in which the Friedmann singularity becomes stable and at the same time
\textit{no supraluminal signals} exist in its vicinity. This possibility was
overlooked in \cite{BNK}.

It is worth to add that also the case $\nu<1/2$ is analyzed in the present
article but it is of no interest since it leads to the strong instability of
the isotropic singularity independently of the question of relativistic causality.

Also it is necessary to stress that here as well as in the old paper
\cite{BNK} only the standard models for a physical fluid is considered for
which the pressure is non-negative and is less than the energy density.

To make the present paper self-contained we will reproduce below the principal
(although updated) points on which the analysis of the work \cite{BNK} was
based. Then to read the present paper there is no necessity to turn to our old publication.

\section{Basic equations in the presence of the shear stresses}

Shear stresses generate an addend $S_{ik}$ to the standard energy-momentum
tensor of a fluid\footnote{We use units in which the Einstein gravitational
constant and the velocity of light are equal to unity. The Greek indices refer
to the three-dimensional space and assume values 1,2,3. The Latin indices
refer to the four-dimensional space-time and take values 0,1,2,3. The time
coordinate is designated as $x^{0}=t$. The interval we write in the old
Landau-Lifschitz fashion:$-ds^{2}=g_{ik}dx^{i}dx^{k}$, where $g_{ik}$ has
signature (-+++). Any time-like vector has negative squared norm. The simple
partial derivatives we designate by comma and covariant derivatives by
semicolon. In synchronous reference system the simple derivatives with respect
to the synchronous time $t$ we denote also by dot.}:
\begin{equation}
T_{ik}=\left(  \varepsilon+p\right)  u_{i}u_{k}+pg_{ik}+S_{ik}\text{ },
\label{P1}%
\end{equation}
and this additional term has to satisfy the following constraints \cite{LL}:%
\begin{equation}
S_{ik}=S_{ki}\text{ },\text{ }S_{k}^{k}=0\text{ },\text{ }u^{i}S_{ik}=0\text{
}. \label{P2}%
\end{equation}
Besides we have the usual normalization condition for the 4-velocity:%
\begin{equation}
u_{i}u^{i}=-1. \label{P3}%
\end{equation}
If the Maxwell's relaxation time $\tau$ of the stresses is not zero then do
not exists any closed expression for $S_{ik}$ in terms of the viscosity
coefficient $\eta$ and 4-gradients of the 4-velocity. Instead the stresses
$S_{ik}$ should be defined from the following differential equations
\cite{Isr1}:
\begin{gather}
S_{ik}+\tau\left(  \delta_{i}^{m}+u_{i}u^{m}\right)  \left(  \delta_{k}%
^{n}+u_{k}u^{n}\right)  S_{mn}{}_{;l}u^{l}=\label{P4}\\
=-\eta\left(  u_{i;k}+u_{k;i}+u^{l}u_{k}u_{i;l}+u^{l}u_{i}u_{k;l}\right)
+\nonumber\\
+\frac{2}{3}\eta\left(  g_{ik}+u_{i}u_{k}\right)  u^{l}{}_{;l}\text{
},\nonumber
\end{gather}
which due to the normalization condition for velocity is compatible with
constraints (\ref{P2}). In case $\tau=0$ expression for $S_{ik},$ following
from this equation, coincides with that one introduced by Landau and Lifschitz
\cite{LL}. If the equations of state $p=p\left(  \varepsilon\right)  ,$
$\eta=\eta\left(  \varepsilon\right)  ,$ $\tau=\tau\left(  \varepsilon\right)
$ are fixed then the Einstein equations%
\begin{equation}
R_{ik}=T_{ik}-\frac{1}{2}g_{ik}T_{l}^{l} \label{P4-1}%
\end{equation}
together with equation (\ref{P4}) for the stresses gives the closed system
where from all quantities of interest, that is $g_{ik},u_{i},\varepsilon
,S_{ik}$ can be found.

Since we are interesting in behaviour of the system in the vicinity to the
cosmological singularity where $\varepsilon\rightarrow\infty$ the viscosity
coefficient $\eta$ in this asymptotic domain can be approximated by the power
law asymptotics%
\begin{equation}
\eta=const\cdot\varepsilon^{\nu}, \label{P5}%
\end{equation}
with some constant exponent $\nu.$ Beforehand the value of this exponent is
unknown then we need to investigate its entire range $-\infty<\nu<\infty.$ As
for the relaxation time $\tau$ the choice is more definite. It is well known
that $\eta/\varepsilon\tau$ represents a measure of velocity of propagation of
the shear excitations. Then we will model this ratio by a positive constant
$f$ (in a more accurate theory $f$ can be a slow varying function on time but
in any case this function should be bounded in order to exclude the appearance
of the supraluminal signals). Consequently we choose the following model for
the relation between relaxation time and viscosity coefficient:%
\begin{equation}
\eta=f\varepsilon\tau,\text{ }f=const. \label{P6}%
\end{equation}
For the dependence $p=p\left(  \varepsilon\right)  $ we follow the standard
approximation with constant parameter $\gamma$:
\begin{equation}
p=\left(  \gamma-1\right)  \varepsilon\text{ },\text{ }1\leqslant\gamma<2.
\label{P7}%
\end{equation}

Now the system (\ref{P1})-(\ref{P7}) is closed and we can search the
asymptotic behaviour of its solution in the vicinity to the cosmological
singularity. It is convenient to work in the synchronous reference system
where the interval is
\begin{equation}
-ds^{2}=-dt^{2}+g_{\alpha\beta}dx^{\alpha}dx^{\beta}. \label{P8}%
\end{equation}
\ 

Our task is to take the standard Friedmann solution in this system as
background and to find the asymptotic (near singularity) solution of the
equations (\ref{P1})-(\ref{P7}) for the linear perturbations around this
background in the same synchronous system. The background solution
is\footnote{It is enough to analyze the flat Friedmann model. As was indicated
in \cite{LK} flatness essentially simplifies calculations and in the same time
the generalization to the closed or open models contributes nothing
principally new to the behavior of the perturbations.}:%
\begin{align}
-ds^{2}  &  =-dt^{2}+R^{2}\left[  \left(  dx^{1}\right)  ^{2}+\left(
dx^{2}\right)  ^{2}+\left(  dx^{3}\right)  ^{2}\right]  ,\text{ }\label{P9}\\
R  &  =\left(  t/t_{c}\right)  ^{2/3\gamma},\text{ }\nonumber
\end{align}

\begin{equation}
\varepsilon=4\left(  3\gamma^{2}t^{2}\right)  ^{-1},\text{ }u_{0}=-1,\text{
}u_{\alpha}=0,\text{ }S_{ik}=0, \label{P10}%
\end{equation}
where $t>0$\ and $t_{c}$\ is some arbitrary positive constant (it is worth to
remark that in the comoving and at the same time synchronous system the right
hand side of the equation (\ref{P4}) is identically zero then the background
value $S_{ik}=0$ indeed satisfies this equation). We have to deal with the
following linear perturbations (as usual any quantity $X$ we write as
$X=X_{\left(  0\right)  }+\delta X$ where $X_{\left(  0\right)  }$ represents
the background value of $X$ ):
\begin{equation}
\delta g_{\alpha\beta},\delta u_{\alpha},\delta\varepsilon,\delta
S_{\alpha\beta}. \label{P11}%
\end{equation}
In the linearized version of the system (\ref{P1})-(\ref{P8}) around the
Friedmann solution (\ref{P9})-(\ref{P10}) will appear only these variations.
The variations $\delta u_{0}$ and $\delta S_{0k}$ can not be of the first
(linear) order because of the exact relations $u_{i}u^{i}=-1$ and $u^{i}%
S_{ik}=0$ and properties (\ref{P10}) of the background. The variations
$\delta\tau$ and $\delta\eta$ of the relaxation time and viscosity
coefficient, although exist as the first order quantities, will disappear from
the linear approximation since they reveal itself only as factors in front of
the terms vanishing for the isotropic Friedmann seed.

Let's introduce for the quantities (\ref{P11}) the following notations:%
\begin{equation}
\delta g_{\alpha\beta}=R^{2}H_{\alpha\beta},\text{ }\delta u_{\alpha
}=V_{\alpha},\text{ }\delta\varepsilon=E,\text{ }\delta S_{\alpha\beta}%
=R^{2}K_{\alpha\beta}. \label{P12}%
\end{equation}

(Here and in the sequel we\textit{\ }use two different entries $\delta
^{\alpha\beta}$\ and $\delta_{\alpha\beta}$ for one and the same 3-dimensional
Kronecker delta). In terms of these quantities the linearized version of the
equations (\ref{P1})-(\ref{P4-1}) in synchronous reference system over the
Friedmann space-time (\ref{P9})-(\ref{P10}) becomes:
\begin{gather}
\ddot{H}_{\alpha\beta}+\frac{3\dot{R}}{R}\dot{H}_{\alpha\beta}+\frac{\dot{R}%
}{R}\delta_{\alpha\beta}\dot{H}+\label{P15}\\
+\frac{1}{R^{2}}\left(  \delta^{\lambda\mu}H_{\alpha\lambda},_{\beta\mu
}+\delta^{\lambda\mu}H_{\beta\lambda},_{\alpha\mu}-\delta^{\lambda\mu
}H_{\alpha\beta},_{\lambda\mu}-H,_{\alpha\beta}\right)  =\nonumber\\
=\left(  2-\gamma\right)  \delta_{\alpha\beta}E+2K_{\alpha\beta}\text{
},\nonumber
\end{gather}%
\begin{equation}
\dot{H},_{\alpha}-\delta^{\lambda\mu}\dot{H}_{\alpha\lambda},_{\mu}%
=2\gamma\varepsilon V_{\alpha}\text{ }, \label{P16}%
\end{equation}%
\begin{equation}
\ddot{H}+\frac{2\dot{R}}{R}\dot{H}=\left(  2-3\gamma\right)  E\text{ },
\label{P17}%
\end{equation}%
\begin{gather}
K_{\alpha\beta}+\tau\dot{K}_{\alpha\beta}=-\frac{\eta}{R^{2}}\left(
V_{\alpha},_{\beta}+V_{\beta},_{\alpha}\right)  +\label{P18}\\
+\frac{2\eta}{3R^{2}}\delta_{\alpha\beta}\delta^{\lambda\mu}V_{\lambda},_{\mu
}-\eta\left(  \dot{H}_{\alpha\beta}-\frac{1}{3}\delta_{\alpha\beta}\dot
{H}\right)  ,\nonumber
\end{gather}
where $H=\delta^{\alpha\beta}H_{\alpha\beta}$. In these formulas $R$ is the
background scale factor given in (\ref{P9}) and $\varepsilon,\tau,\eta$ are
the background values of these quantities defined by the the relations
(\ref{P5}), (\ref{P6}) and (\ref{P10}) (in principle they should be written as
$\varepsilon_{\left(  0\right)  },\tau_{\left(  0\right)  },\eta_{\left(
0\right)  }$ but we omit the index $\left(  0\right)  $ to simplify the writing).

To find the general solution of these equations we apply the technique
invented by Lifschitz \cite{L1} (see also \cite{LK}) and used by him to
analyze the stability of the Friedmann solution for the perfect liquid. Since
all coefficients in the differential equations (\ref{P15})-(\ref{P18}) do not
depend on spatial coordinates we can represent all quantities of interest in
the form of the 3-dimensional Fourier integrals to reduce these equations to
the system of the ordinary differential equations in time for the
corresponding Fourier coefficients. First of all we substitute the expression
for $E$ from equation (\ref{P17}) to the right hand side of equation
(\ref{P15}) and expression for $V_{\alpha}$ from (\ref{P16}) to the right hand
side of (\ref{P18}). This gives the closed system of equations for tensorial
perturbations $H_{\alpha\beta}$ and $K_{\alpha\beta}$ and corresponding system
of ordinary differential equations in time for their Fourier coefficients
$\tilde{H}_{\alpha\beta}$ and $\tilde{K}_{\alpha\beta}$ (any space-time field
$\Phi\left(  t,x^{1},x^{2},x^{3}\right)  $ we represent as $\Phi\left(
t,x^{1},x^{2},x^{3}\right)  =%
{\displaystyle\int}
\tilde{\Phi}\left(  t,k_{1},k_{2},k_{3}\right)  e^{ik_{\alpha}x^{\alpha}}%
d^{3}k$). Any symmetric tensorial Fourier coefficient containing six
independent components can be expended in the Lifshitz basis which consists of
the six basic elements. These elements can be constructed from an orthonormal
triad $\left(  l_{\alpha}^{\left(  1\right)  },l_{\alpha}^{\left(  2\right)
},l_{\alpha}^{\left(  3\right)  }\right)  $ in the euclidean $k$-space where%

\begin{equation}
l_{\alpha}^{\left(  1\right)  }=k_{\alpha}/k\text{ },\text{ }k=\sqrt
{\delta^{\alpha\beta}k_{\alpha}k_{\beta}}), \label{P14-2}%
\end{equation}
that is $l_{\alpha}^{\left(  1\right)  }$ is the unit directional vector of $k
$-space and\textbf{\ }$l_{\alpha}^{\left(  2\right)  },l_{\alpha}^{\left(
3\right)  }$ are another two unit vectors normal to $k_{\alpha}$ and to each
other. The aforementioned basic elements are:%

\begin{equation}
Q_{\alpha\beta}=\frac{1}{3}\delta_{\alpha\beta}\text{ },\text{ \ }%
P_{\alpha\beta}=\frac{1}{3}\delta_{\alpha\beta}-\frac{k_{\alpha}k_{\beta}%
}{k^{2}}, \label{P14-3}%
\end{equation}%
\begin{equation}
L_{\alpha\beta}^{\left(  2\right)  }=\frac{k_{\alpha}}{k}l_{\beta}^{\left(
2\right)  }+\frac{k_{\beta}}{k}l_{\alpha}^{\left(  2\right)  },\text{
}L_{\alpha\beta}^{\left(  3\right)  }=\frac{k_{\alpha}}{k}l_{\beta}^{\left(
3\right)  }+\frac{k_{\beta}}{k}l_{\alpha}^{\left(  3\right)  }, \label{P14-4}%
\end{equation}%
\begin{equation}
G_{\alpha\beta}^{\left(  2\right)  }=l_{\alpha}^{\left(  2\right)  }l_{\beta
}^{\left(  3\right)  }+l_{\beta}^{\left(  2\right)  }l_{\alpha}^{\left(
3\right)  },\text{ }G_{\alpha\beta}^{\left(  3\right)  }=l_{\alpha}^{\left(
2\right)  }l_{\beta}^{\left(  2\right)  }-l_{\alpha}^{\left(  3\right)
}l_{\beta}^{\left(  3\right)  }. \label{P14-5}%
\end{equation}
Then $\tilde{H}_{\alpha\beta}$ and $\tilde{K}_{\alpha\beta}$ can be expended
in the following way:%
\begin{equation}
\tilde{H}_{\alpha\beta}=\lambda P_{\alpha\beta}\text{ }+\mu Q_{\alpha\beta}+%
{\displaystyle\sum\limits_{J=2}^{3}}
\left[  \sigma_{\left(  J\right)  }L_{\alpha\beta}^{\left(  J\right)  }%
+\omega_{\left(  J\right)  }G_{\alpha\beta}^{\left(  J\right)  }\right]  ,
\label{P13}%
\end{equation}%
\begin{equation}
\tilde{K}_{\alpha\beta}=AP_{\alpha\beta}\text{ }+%
{\displaystyle\sum\limits_{J=2}^{3}}
\left[  B_{\left(  J\right)  }L_{\alpha\beta}^{\left(  J\right)  }+D_{\left(
J\right)  }G_{\alpha\beta}^{\left(  J\right)  }\right]  , \label{P14}%
\end{equation}
(here we introduced the new index $J$ which takes only two values $2$ and
$3$), where the amplitudes $\lambda,\mu,\sigma_{\left(  J\right)  }%
,\omega_{\left(  J\right)  },A,B_{\left(  J\right)  },D_{\left(  J\right)  }$
depend on time (and on the components of the wave vector).

Only $Q_{\alpha\beta}$ has non-zero contraction $\delta^{\alpha\beta}%
Q_{\alpha\beta}$ , that's why in the expansion (\ref{P14}) for the shear
stresses this component is absent (remember that the second condition of
(\ref{P2}) calls $\delta^{\alpha\beta}K_{\alpha\beta}=0$). The reason why the
Lifshitz basis is better than any other lies in the fact that in this basis
the system of the differential equations (\ref{P15})-(\ref{P18}) re-written in
terms of the $\lambda,\mu,\sigma_{\left(  J\right)  },\omega_{\left(
J\right)  },A,B_{\left(  J\right)  },D_{\left(  J\right)  }$ splits in the
three separate and independent subsets: the first for $\lambda,\mu,A,$ the
second for $\sigma_{\left(  J\right)  },B_{\left(  J\right)  }$ and the third
for $\omega_{\left(  J\right)  },D_{\left(  J\right)  }.$ The equations for
$\lambda,\mu,A$ are:%

\begin{equation}
\ddot{\mu}+\frac{3\gamma\dot{R}}{R}\dot{\mu}+\frac{k^{2}(3\gamma-2)}{3R^{2}%
}\left(  \lambda+\mu\right)  =0\text{ }, \label{P20}%
\end{equation}%
\begin{equation}
\ddot{\lambda}+\frac{3\dot{R}}{R}\dot{\lambda}-\frac{k^{2}}{3R^{2}}\left(
\lambda+\mu\right)  =2A\text{ }, \label{P21}%
\end{equation}%
\begin{equation}
\text{\ }A+\tau\dot{A}=-\eta\dot{\lambda}-\frac{2\eta k^{2}}{3\gamma
\varepsilon R^{2}}\left(  \dot{\lambda}+\dot{\mu}\right)  \text{ }.
\label{P22}%
\end{equation}
For the four amplitudes $\sigma_{\left(  J\right)  },B_{\left(  J\right)  }$
we have:%
\begin{equation}
\ddot{\sigma}_{\left(  J\right)  }+\frac{3\dot{R}}{R}\dot{\sigma}_{\left(
J\right)  }=2B_{\left(  J\right)  }\text{ },\text{ \ } \label{P23}%
\end{equation}%
\begin{equation}
B_{\left(  J\right)  }+\tau\dot{B}_{\left(  J\right)  }=-\eta\dot{\sigma
}_{\left(  J\right)  }-\frac{\eta k^{2}}{2\gamma\varepsilon R^{2}}\dot{\sigma
}_{\left(  J\right)  }\text{ }, \label{P24}%
\end{equation}

and equations for two pairs $\omega_{\left(  J\right)  },D_{\left(  J\right)
}$ are:%
\begin{equation}
\ddot{\omega}_{\left(  J\right)  }+\frac{3\dot{R}}{R}\dot{\omega}_{\left(
J\right)  }+\frac{k^{2}}{R^{2}}\omega_{\left(  J\right)  }=2D_{\left(
J\right)  }\text{ }, \label{P25}%
\end{equation}%
\begin{equation}
D_{\left(  J\right)  }+\tau\dot{D}_{\left(  J\right)  }=-\eta\dot{\omega
}_{\left(  J\right)  }\text{ }. \label{P25-A}%
\end{equation}

If we know functions $\lambda,\mu$ and $\sigma_{\left(  J\right)  }$ the
Fourier components $\tilde{V}_{\alpha},\tilde{E}$ of perturbations of velocity
and energy density, as follows from the equations (\ref{P16}) and (\ref{P17})
(also making use the equation (\ref{P20}) to eliminate the second derivative
$\ddot{\mu}$), can be expressed in terms of these functions by the relations:%
\begin{equation}
\tilde{V}_{\alpha}=\frac{ik}{2\gamma\varepsilon}\left[  \frac{2}{3}\left(
\dot{\lambda}+\dot{\mu}\right)  \frac{k_{\alpha}}{k}-%
{\displaystyle\sum\limits_{J=2}^{3}}
\dot{\sigma}_{\left(  J\right)  }l_{\alpha}^{\left(  J\right)  }\right]  ,
\label{P26}%
\end{equation}%
\begin{equation}
\tilde{E}=\frac{\dot{R}}{R}\dot{\mu}+\frac{k^{2}}{3R^{2}}\left(  \lambda
+\mu\right)  \text{ }. \label{P27}%
\end{equation}

\section{On the propagation of the short wavelength pulses}

To study the waves of the short wavelength (formally $k\rightarrow\infty$) it
is convenient to pass to the conformally flat version of the Friedman metric
$-ds^{2}=R^{2}\left(  T\right)  \left[  -dT^{2}+\left(  dx^{1}\right)
^{2}+\left(  dx^{2}\right)  ^{2}+\left(  dx^{3}\right)  ^{2}\right]  $,
introducing the new time variable $T$ by the relation $dT=dt/R.$ Then in the
limit when $k$ dominates in the equations (\ref{P20})-(\ref{P25-A}) this
system has the following set of solutions:
\begin{gather}
\lambda=\lambda^{sva}\left(  T\right)  e^{i\upsilon_{1}kT},\text{ }\mu
=\mu^{sva}\left(  T\right)  e^{i\upsilon_{1}kT},\text{ }\label{W1}\\
A=A^{sva}\left(  T\right)  e^{i\upsilon_{1}kT},\nonumber
\end{gather}%
\begin{equation}
\sigma_{\left(  J\right)  }=\sigma_{\left(  J\right)  }^{sva}\left(  T\right)
e^{i\upsilon_{2}kT},\text{ }B_{\left(  J\right)  }=B_{\left(  J\right)
}^{sva}\left(  T\right)  e^{i\upsilon_{2}kT}, \label{W2}%
\end{equation}%
\begin{equation}
\omega_{\left(  J\right)  }=\omega_{\left(  J\right)  }^{sva}\left(  T\right)
e^{i\upsilon_{3}kT},\text{ }D_{\left(  J\right)  }=D_{\left(  J\right)
}^{sva}\left(  T\right)  e^{i\upsilon_{3}kT}, \label{W3}%
\end{equation}
with large phases and slow varying amplitudes (index $sva$). Substituting
these expressions into the equations (\ref{P20})-(\ref{P25-A}) and keeping
only the terms of highest order with respect to the large quantity $k$ one get
the velocities of propagation of perturbations:%
\begin{equation}
\upsilon_{1}^{2}=\gamma-1+\frac{4f}{3\gamma},\text{ }\upsilon_{2}^{2}=\frac
{f}{\gamma},\text{ }\upsilon_{3}^{2}=1. \label{W4}%
\end{equation}
This result have been obtained in \cite{BNK} and it shows that gravitational
waves (perturbation $\omega_{\left(  J\right)  },D_{\left(  J\right)  }$)
propagate with velocity of light but in order to exclude the supraluminal
signals for two other types of perturbations it is necessary to demand
$\upsilon_{1}^{2}<1$ and $\upsilon_{2}^{2}<1.$ Both of these conditions in the
region $1\leqslant\gamma<2$ will be satisfied if
\begin{equation}
\text{ }f<\frac{3}{4}\gamma\left(  2-\gamma\right)  . \label{W5}%
\end{equation}

\section{Extreme vicinity to the singularity}

The useful property of the equations (\ref{P20})-(\ref{P25-A}) is an essential
simplification and unification of their mathematical forms near singularity.
Indeed near the singular point $t\rightarrow0$ in the limit when $t$ is much
less than everything else (including $t\ll k^{-1})$ we can neglect in these
equations by all terms containing the factor $k^{2}R^{-2\text{ }}$which are
much smaller than all other terms\footnote{This is because in such region
$k^{2}R^{-2}\ll t^{-2}$ (we remind that $R^{2}\sim t^{4/3\gamma}$ and
$4/3\gamma<2$ for $\gamma\geqslant1$) then the terms containing the time
derivatives in equations (\ref{P20}), (\ref{P21}) and (\ref{P25}) are much
greater than those containing the factor $k^{2}R^{-2}$ (the evaluation of the
order of magnitude of the derivatives of the functions near singularity
$t\rightarrow0$ follow the rule $\dot{\lambda}\sim t^{-1}\lambda$ and its
validity can be checked directly by the resulting solution the derivation of
which is based on this rule). Also in the equations (\ref{P22}) and
(\ref{P24}) the terms containing factor $k^{2}\varepsilon^{-1}R^{-2}$ can be
neglected because this factor has order $t^{2-4/3\gamma}$ and tends to zero
when $t\rightarrow0.$}. Consequently the asymptotic form of the equations
(\ref{P20})-(\ref{P25-A}) in the vicinity to singularity is:
\begin{equation}
\ddot{\mu}+\frac{2}{t}\dot{\mu}=0, \label{P33}%
\end{equation}%
\begin{equation}
\ddot{\lambda}+\frac{2}{\gamma t}\dot{\lambda}=2A,\text{ \ }A+\tau\dot
{A}=-\eta\dot{\lambda}. \label{P34}%
\end{equation}%
\begin{equation}
\ddot{\sigma}_{\left(  J\right)  }+\frac{2}{\gamma t}\dot{\sigma}_{\left(
J\right)  }=2B_{\left(  J\right)  },\text{ \ }B_{\left(  J\right)  }+\tau
\dot{B}_{\left(  J\right)  }=-\eta\dot{\sigma}_{\left(  J\right)  }.
\label{P35}%
\end{equation}%
\begin{equation}
\ddot{\omega}_{\left(  J\right)  }+\frac{2}{\gamma t}\dot{\omega}_{\left(
J\right)  }=2D_{\left(  J\right)  },\text{ \ }D_{\left(  J\right)  }+\tau
\dot{D}_{\left(  J\right)  }=-\eta\dot{\omega}_{\left(  J\right)  }.
\label{P36}%
\end{equation}
In the solution $\mu=C_{\mu}^{\left(  -1\right)  }t^{-1}+$ $C_{\mu}^{\left(
0\right)  }$ of the first equation both arbitrary constants $C_{\mu}^{\left(
-1\right)  }$ and $C_{\mu}^{\left(  0\right)  }$can be removed by the
coordinate transformations which still exist in the synchronous system
\cite{LK}, that is $\mu$ in this approximation represents a pure gauge (non
physical) excitation. We can take%
\begin{equation}
\mu=C_{\mu}^{\left(  0\right)  } \label{P36-2}%
\end{equation}
without loss of generality but keeping in mind that also constant $C_{\mu
}^{\left(  0\right)  }$ can be put to zero\footnote{The appearance of the
arbitrary gauge perturbation $C_{\mu}^{\left(  -1\right)  }\left(  k_{1}%
,k_{2},k_{3}\right)  $ corresponds to the change under which the cosmological
singularity becomes non-simultaneous that is instead of $t=0$ it take the
equation $t-\varphi\left(  x^{1},x^{2},x^{3}\right)  =0,$ where $\varphi
=-\frac{\gamma}{4}\int C_{\mu}^{\left(  -1\right)  }e^{ik_{\alpha}x^{\alpha}%
}d^{3}k$. Elimination of the constant $C_{\mu}^{\left(  -1\right)  }$
corresponds to that choice of the initial hypersurface in the synchronous
system for which Friedmann singularity remains simultaneous independently of
the presence of inhomogeneous perturbations. After fixing this gauge the
remaining non-physical degrees of freedom in the synchronous system correspond
to the 3-dimensional coordinate transformations $x^{\alpha}=x^{\alpha}\left(
\acute{x}^{1},\acute{x}^{2},\acute{x}^{3}\right)  $ by which we can eliminate
the three arbitrary parameters $C_{\mu}^{\left(  0\right)  }$ and
$C_{\sigma_{\left(  J\right)  }}^{\left(  0\right)  }$ (these last two will
appear later).}. The other pairs of perturbations are described by the
identical equations so it is enough to consider only one such pair, for
example $(\lambda,$ $A).$ As analysis shows there are three principally
different characters of behaviour of perturbations for the three different
ranges of values of the index $\nu$ in formula (\ref{P5}) for the viscosity
coefficient, namely $\nu<1/2,$ $\nu>1/2$ and $\nu=1/2.$ For the first two
ranges it is convenient to represent relations (\ref{P5}) and (\ref{P6})
(taking into account expression (\ref{P10}) for $\varepsilon$) in the
following form:%
\begin{equation}
\frac{\eta}{\tau}=\frac{4f}{3\gamma^{2}t^{2}},\text{ }\tau=\left(  \frac
{t}{t_{\tau}}\right)  ^{1-2\nu}t,\text{ } \label{P37}%
\end{equation}
with some arbitrary positive constant $t_{\tau}.$ Then from the equations
(\ref{P34}) follow that $\dot{\lambda}$ and $A$ can be expressed in term of an
auxiliary function $F\left(  t\right)  $ as:
\begin{equation}
\dot{\lambda}=2F,\text{ }A=\dot{F}+\frac{2}{\gamma t}F, \label{P38}%
\end{equation}
after which equations (\ref{P34}) reduce to one ordinary equation of second
order for $F$:
\begin{gather}
\ddot{F}+\frac{1}{t}\left[  \frac{2}{\gamma}+\left(  \frac{t}{t_{\tau}%
}\right)  ^{2\nu-1}\right]  \dot{F}+\label{P39}\\
+\frac{1}{t^{2}}\left[  -\frac{2}{\gamma}+\frac{2}{\gamma}\left(  \frac
{t}{t_{\tau}}\right)  ^{2\nu-1}+\frac{8f}{3\gamma^{2}}\right]  F=0\nonumber
\end{gather}
If instead of $t$ and $F\left(  t\right)  $ we introduce the new time variable
$y$ and new function $W\left(  y\right)  $ by the relations:%
\begin{align}
y  &  =\frac{1}{\left\vert 2\nu-1\right\vert }\left(  \frac{t}{t_{\tau}%
}\right)  ^{2\nu-1},\text{ \ }\label{P40}\\
F  &  =\left\vert 2\nu-1\right\vert ^{\alpha}y^{\alpha}\exp\left(
-\frac{\left\vert 2\nu-1\right\vert y}{2\left(  2\nu-1\right)  }\right)
W\left(  y\right)  ,\nonumber
\end{align}
where
\begin{equation}
\alpha=\frac{\gamma\left(  1-\nu\right)  -1}{\gamma\left(  2\nu-1\right)  },
\label{P41}%
\end{equation}
then (\ref{P39}) gives the Whittaker equation \cite{Gr-Ry}:%
\begin{equation}
W_{,\text{ }yy}+\left(  -\frac{1}{4}+\frac{L}{y}+\frac{1-4M^{2}}{4y^{2}%
}\right)  W=0, \label{P42}%
\end{equation}
where constants $L$ and $M$ are:%
\begin{equation}
L=\frac{\gamma(1-\nu)+1}{\gamma\left\vert 2\nu-1\right\vert },\text{
\ }M=\sqrt{\frac{3\left(  \gamma+2\right)  ^{2}-32f}{12\gamma^{2}\left(
2\nu-1\right)  ^{2}}},\text{ \ } \label{P43}%
\end{equation}
It is easy to check that due to the condition of causality (\ref{W5}) the
quantity $3\left(  \gamma+2\right)  ^{2}-32f$ under the square root in
expression for $M$ can never be negative. Then $M$ is real and without loss of
generality we can choose its positive branch $M>0.$

For the boundary value $\nu=1/2$ the representation (\ref{P37}) and
(\ref{P39})-(\ref{P43}) does not works and this special case we will consider
separately (see below).

\subsection{\textit{Case }$\nu<1/2.$}

In this case, as follows from (\ref{P40}), near singularity ($t\rightarrow0$)
the variable $y\rightarrow\infty$. Then the asymptotic behaviour of the
function $W\left(  y\right)  $ at infinity is characterized by the
superposition of two terms $y^{-L}e^{y/2}$ and $y^{L}e^{-y/2}$ (this can be
seen directly from the equation (\ref{P42}) without necessity to go to a
reference book for the asymptotic properties of the two Whittaker fundamental
solutions $W_{L,M}\left(  y\right)  $ and $W_{-L,M}\left(  -y\right)  $). Then
relations (\ref{P40}) and (\ref{P38}) show that perturbations contain the
strongly divergent mode for which $\lambda,A\sim\exp\left[  \frac{1}{2\left(
1-2\nu\right)  }\left(  \frac{t}{t_{\tau}}\right)  ^{2\nu-1}\right]  .$ This
mode destroys the background regime. Consequently the values $\nu<1/2$ are of
no interest for us since in this case does not exists a general solution of
the gravitational equations with the Friedmann singularity.

\subsection{Case $\nu>1/2.$}

For $\nu>1/2$ the singularity $t\rightarrow0$ corresponds to $y\rightarrow0.$
In this asymptotic region the superposition of two modes $W_{\pm}\sim
y^{\frac{1}{2}\pm M}$ forms the general solution for the Whittaker
equation\footnote{We can ignore the particular cases when $2M$ takes the
integer values $2M=1,2,...$ (the value $M=0$ is automatically impossible since
for $1\leqslant\gamma<2$ it violate the causality condition (\ref{W5})). The
subtleties with the second Whittaker mode for the integer value of $2M$ (so
called "logarithmic case") do not introduce the principal changes for the
behaviour of perturbations near singularity.}. For the function $F\left(
t\right)  $ this corresponds to the modes $F_{\pm}\sim t^{\left(
2\nu-1\right)  \left(  \alpha+\frac{1}{2}\pm M\right)  }$ which gives the
following asymptotic behaviour for two time-dependent perturbation modes for
metric: $\lambda_{\pm}\sim t^{\left(  2\nu-1\right)  \left(  \alpha+\frac
{1}{2}\pm M\right)  +1}.$ With definitions (\ref{P41}) and (\ref{P43}) for
constants $\alpha$ and $M$ we have:%
\begin{equation}
\lambda_{\pm}\sim t^{s_{\pm}},\text{ \ }s_{\pm}=\frac{3\gamma-2}{2\gamma}%
\pm\sqrt{\frac{3\left(  \gamma+2\right)  ^{2}-32f}{12\gamma^{2}}}\text{ },
\label{P44}%
\end{equation}
For stability of the Friedmann solution it is necessary for both exponents
$s_{\pm}$ to be positive ($\lambda_{\pm}$ must disappear in the limit
$t\rightarrow0$). Because the first term in $s_{\pm}$ in the region
$1\leqslant\gamma<2$ is positive and the square root is positive we need to
provide only the inequality $s_{-}>0$ and it is easy to show that this is
equivalent to the restriction $f>\frac{3}{4}\gamma\left(  2-\gamma\right)  $
for the constant $f.$ However, this restriction is \textit{exactly opposite to
the causality condition} (\ref{W5}). Consequently, also for $\nu>1/2,$
assuming the absence of the supraluminal excitations, there is no way to
provide stability of the Friedmann solution near singularity. This result has
been obtained already in \cite{BNK}. It is worth to remark that this state of
affairs is in conformity with the general statement, already expressed in the
literature, on the connection between the existence of the supraluminal
signals and instability of the equilibrium states \cite{GL,His}.

\subsection{Case $\nu=1/2.$}

\begin{figure}
[ptb]
\begin{center}
\includegraphics[
width=\columnwidth
]%
{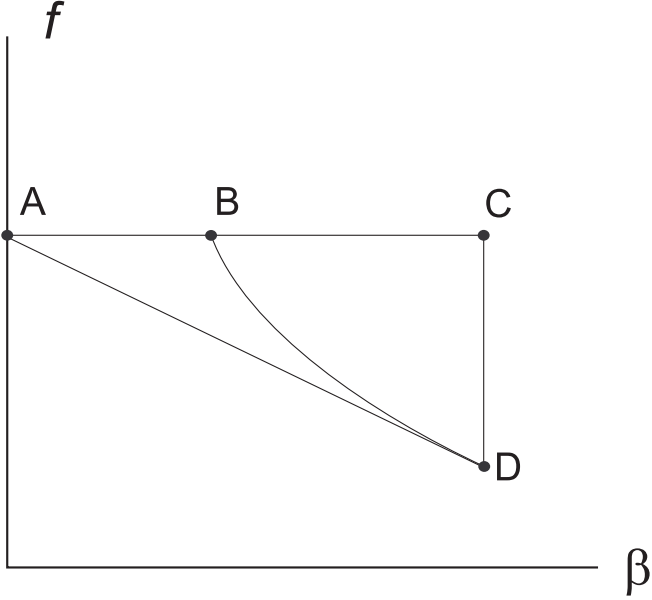}
\caption{Two regions ($ABD$ and $BCD$) in the plane of parameters $(f,\beta)$
where the Friedmann singularity is stable are shown. For each fixed value of
parameter $\gamma$ the coordinates of the points $A,B,C,D$ are fixed and all
acceptable values of $f,\beta$ for each $\gamma$ are located in the region
$ABCD.$ In $ABD$ both exponents $s_{1}$ and $s_{2}$ are real and positive and
at the same time no supraluminal velocities exists. In $BCD$ exponents $s_{1}$
and $s_{2}$ are complex conjugated to each other but with the positive real
part and again no supraluminal signals exist. The coordinates of the boundary
points $A,B,C,D$ depend on the constant $\gamma$ (which has been chosen from
the region $1\leqslant\gamma<2$) and they are (the first coordinate indicate
the value of $f$ and the second \ the value of $\beta)$: $A[\frac{3}{4}%
\gamma(2-\gamma),$ $0],$ $B[\frac{3}{4}\gamma(2-\gamma),\frac{2+\gamma
-\sqrt{8\gamma(2-\gamma)}}{\gamma}],$ $C[\frac{3}{4}\gamma(2-\gamma),$
$\frac{3\gamma-2}{\gamma}],$ $D[\frac{3}{8}\left(  2-\gamma\right)  ^{2},$
$\frac{3\gamma-2}{\gamma}].$ The straight line $AD$ has the equation
$f=\frac{3}{8}\gamma\left(  2-\gamma\right)  (2-\beta).$ The curve $BD$ is the
piece of parabola $f=$ $\frac{3}{32}\left[  2-\gamma\left(  \beta-1\right)
\right]  ^{2}.$}%
\label{fig.1}%
\end{center}
\end{figure}

For $\nu=1/2$ instead of (\ref{P37}) we have to write:
\begin{equation}
\frac{\eta}{\tau}=\frac{4f}{3\gamma^{2}t^{2}},\text{ }\tau=\frac{t}{\beta
},\text{ } \label{P45}%
\end{equation}
where $\beta$ is some dimensionless positive arbitrary constant. Relations
(\ref{P38}) are the same and equation for the auxiliary function $F\left(
t\right)  $, which follows from (\ref{P34}) and (\ref{P45}) becomes:%
\begin{equation}
\ddot{F}+\frac{1}{t}\left(  \frac{2}{\gamma}+\beta\right)  \dot{F}+\frac
{1}{t^{2}}\left(  -\frac{2}{\gamma}+\frac{2\beta}{\gamma}+\frac{8f}%
{3\gamma^{2}}\right)  F=0. \label{P46}%
\end{equation}
This equation has exact solutions in the form of two power law modes with
power exponents following from the corresponding quadratic equation. Using the
relation $\dot{\lambda}=2F$ it is easy to show that the final result for the
perturbation $\lambda$ is:%
\begin{equation}
\lambda=C_{\lambda}^{\left(  0\right)  }+C_{\lambda}^{\left(  1\right)
}t^{s_{1}}+C_{\lambda}^{\left(  2\right)  }t^{s_{2}}, \label{P47}%
\end{equation}
where $C_{\lambda}^{\left(  0\right)  },C_{\lambda}^{\left(  1\right)
},C_{\lambda}^{\left(  2\right)  }$ are three arbitrary constants (depending
on the wave vector) and%
\begin{equation}
s_{1,2}=\frac{3\gamma-\gamma\beta-2}{2\gamma}\pm\frac{1}{2\gamma}\sqrt{\left(
\gamma-\gamma\beta+2\right)  ^{2}-\frac{32f}{3}}, \label{P48}%
\end{equation}
where the sign plus corresponds to $s_{1}$ and minus to $s_{2}$ and square
root we take to be positive in case when it is real. In order to have
$\lambda\ll1$ at $t\rightarrow0$ both exponents $s_{1}$ and $s_{2}$ should be
either positive or they should \ have the positive real part. At the same time
in both cases we have to satisfy the relativistic causality condition
(\ref{W5}). Then we have two possibilities. Either
\begin{gather}
3\gamma-\gamma\beta-2>0,\text{ }\left(  \gamma-\gamma\beta+2\right)
^{2}-\frac{32f}{3}>0,\text{ \ }\label{P49}\\
f<\frac{3}{4}\gamma\left(  2-\gamma\right)  ,\text{ }3\gamma-\gamma
\beta-2>\sqrt{\left(  \gamma-\gamma\beta+2\right)  ^{2}-\frac{32f}{3}%
},\nonumber
\end{gather}
in which case $s_{1}$ and $s_{2}$ are real and positive, or%
\begin{align}
3\gamma-\gamma\beta-2  &  >0,\text{ \ }\left(  \gamma-\gamma\beta+2\right)
^{2}-\frac{32f}{3}<0,\text{ \ }\label{P50}\\
f  &  <\frac{3}{4}\gamma\left(  2-\gamma\right)  ,\nonumber
\end{align}
which corresponds to the complex conjugated $s_{1}$ and $s_{2}$ but with
positive real part. It is easy to show that in the space of parameters
$f,\beta,\gamma$ there are two regions, exposed on the Fig. \ref{fig.1}, in
which either the first or the second of these sets of requirements is
satisfied. The set of inequalities (\ref{P49}) is satisfied in the triangle
$ABD$ and the set (\ref{P50}) is valid in the triangle $BCD$. The caption to
this figure contains all necessary information on the admissible domains for
the values of the parameters $f,\beta,\gamma$.%

Fom the first of the equations (\ref{P34}) follows amplitude $A$:%
\begin{equation}
A=q_{1}C_{\lambda}^{\left(  1\right)  }t^{s_{1}-2}+q_{2}C_{\lambda}^{\left(
2\right)  }t^{s_{2}-2}, \label{P51}%
\end{equation}
where%
\begin{equation}
q_{1}=\frac{s_{1}\left(  \gamma s_{1}-\gamma+2\right)  }{2\gamma}\text{
},\text{ \ }q_{2}=\frac{s_{2}\left(  \gamma s_{2}-\gamma+2\right)  }{2\gamma
}\text{ }. \label{P52}%
\end{equation}

Now one can make asymptotics for $\lambda,\mu$ and $A$ more precise taking
into account those terms in equations (\ref{P20})- (\ref{P22}) containing the
factor $k^{2}R^{-2}$, that is the terms which have been neglected in the first
approximation. Analysis shows that their influence consists in generation the
small time-dependent corrections to the arbitrary constants appeared in the
first approximation. These corrections are completely expressible in terms of
parameters of the first approximation, that is they do not bring any new
arbitrariness in the solution. The fact is that the exact general solution of
equations (\ref{P20})- (\ref{P22}) for the functions $\lambda,\mu$ and
$t^{2}A$ has the form $F^{\left(  0\right)  }\left(  t\right)  +t^{s_{1}%
}F^{\left(  1\right)  }\left(  t\right)  +t^{s_{2}}F^{\left(  2\right)
}\left(  t\right)  $ with the same exponents $s_{1}$ and $s_{2}$ given by the
formula (\ref{P48}) and with functions $F^{\left(  0\right)  },F^{\left(
1\right)  },F^{\left(  2\right)  }$ each of which can be expressed in the form
of the Taylor series in the small parameter $\zeta$:%
\begin{equation}
\zeta=\left(  kt/R\right)  ^{2}, \label{P52-1}%
\end{equation}
which parameter tends to zero in the limit $t\rightarrow0$ because $\left(
kt/R\right)  ^{2}=k^{2}t_{c}^{4/3\gamma}t^{2\left(  3\gamma-2\right)
/3\gamma}$ and $\gamma\geqslant1$. The first time-independent terms in these
power series are just the arbitrary constants figured in the formulas
(\ref{P36-2}),(\ref{P47}) and (\ref{P51}). The general structure of the exact
solution for $\lambda,\mu$ and $t^{2}A$ is:%
\begin{align}
\lambda &  =\left(  C_{\lambda}^{\left(  0\right)  }+\alpha_{\lambda
1}^{\left(  0\right)  }\zeta+\alpha_{\lambda2}^{\left(  0\right)  }\zeta
^{2}...\right)  +\label{P53}\\
&  +\left(  C_{\lambda}^{\left(  1\right)  }+\alpha_{\lambda1}^{\left(
1\right)  }\zeta+\alpha_{\lambda2}^{\left(  1\right)  }\zeta^{2}...\right)
t^{s_{1}}+\nonumber\\
&  +\left(  C_{\lambda}^{\left(  2\right)  }+\alpha_{\lambda1}^{\left(
2\right)  }\zeta+\alpha_{\lambda2}^{\left(  2\right)  }\zeta^{2}...\right)
t^{s_{2}},\nonumber
\end{align}%
\begin{align}
\mu &  =\left(  C_{\mu}^{\left(  0\right)  }+\alpha_{\mu1}^{\left(  0\right)
}\zeta+\alpha_{\mu2}^{\left(  0\right)  }\zeta^{2}...\right)  +\label{P54}\\
&  +\left(  \alpha_{\mu1}^{\left(  1\right)  }\zeta+\alpha_{\mu2}^{\left(
1\right)  }\zeta^{2}...\right)  t^{s_{1}}+\nonumber\\
&  +\left(  \alpha_{\mu1}^{\left(  2\right)  }\zeta+\alpha_{\mu2}^{\left(
2\right)  }\zeta^{2}...\right)  t^{s_{2}},\nonumber
\end{align}%
\begin{align}
t^{2}A  &  =\left(  \alpha_{A1}^{\left(  0\right)  }\zeta+\alpha_{A2}^{\left(
0\right)  }\zeta^{2}...\right)  +\label{P55}\\
&  +\left(  q_{1}C_{\lambda}^{\left(  1\right)  }+\alpha_{A1}^{\left(
1\right)  }\zeta+\alpha_{A2}^{\left(  1\right)  }\zeta^{2}...\right)
t^{s_{1}}+\nonumber\\
&  +\left(  q_{2}C_{\lambda}^{\left(  2\right)  }+\alpha_{A1}^{\left(
2\right)  }\zeta+\alpha_{A2}^{\left(  2\right)  }\zeta^{2}...\right)
t^{s_{2}},\nonumber
\end{align}
where all $\alpha$-coefficients in front of the powers of parameter $\zeta$
are constant quantities which depend on the four arbitrary constants $C_{\mu
}^{\left(  0\right)  },C_{\lambda}^{\left(  0\right)  },C_{\lambda}^{\left(
1\right)  },C_{\lambda}^{\left(  2\right)  }$ and external numbers
$f,\beta,\gamma$. There is no big sense in taking into account corrections
containing the powers of $\zeta$ in the factors in front of the powers
$t^{s_{1}}$ and $t^{s_{2}}$ since this would give the small unimportant
addends to the asymptotics. The same is true for the corrections of the orders
$\zeta^{2}$ and higher in terms which do not contain powers $t^{s_{1}}$ and
$t^{s_{2}}$. However, to keep the terms $\alpha_{\lambda1}^{\left(  0\right)
}\zeta$ , $\alpha_{\mu1}^{\left(  0\right)  }\zeta$ and $\alpha_{A1}^{\left(
0\right)  }\zeta$ in the asymptotics is necessary because in general they,
although small, play a role in the behaviour of the solution and the first
non-vanishing term in the asymptotic expression for the energy density depends
on them. Calculations gives the following result for the coefficients
$\alpha_{\lambda1}^{\left(  0\right)  },\alpha_{\mu1}^{\left(  0\right)  }$
and $\alpha_{A1}^{\left(  0\right)  }$:%
\begin{equation}
a_{\lambda1}^{\left(  0\right)  }=\frac{3\gamma^{2}\left(  3\gamma
\beta-4\right)  }{2\left(  3\gamma-2\right)  \left[  24f+\left(
3\gamma+2\right)  \left(  3\gamma\beta-4\right)  \right]  }\left(  C_{\lambda
}^{\left(  0\right)  }+C_{\mu}^{\left(  0\right)  }\right)  , \label{P55-1}%
\end{equation}%
\begin{equation}
a_{\mu1}^{\left(  0\right)  }=-\frac{3\gamma^{2}}{2\left(  9\gamma-4\right)
}\left(  C_{\lambda}^{\left(  0\right)  }+C_{\mu}^{\left(  0\right)  }\right)
, \label{P55-2}%
\end{equation}%
\begin{equation}
a_{A1}^{\left(  0\right)  }=-\frac{4f}{24f+\left(  3\gamma+2\right)  \left(
3\gamma\beta-4\right)  }\left(  C_{\lambda}^{\left(  0\right)  }+C_{\mu
}^{\left(  0\right)  }\right)  . \label{P55-3}%
\end{equation}
Then the final sufficient asymptotics for the amplitudes $\lambda,\mu$ and $A$
is:%
\begin{equation}
\lambda=C_{\lambda}^{\left(  0\right)  }+C_{\lambda}^{\left(  1\right)
}t^{s_{1}}+C_{\lambda}^{\left(  2\right)  }t^{s_{2}}+\alpha_{\lambda
1}^{\left(  0\right)  }k^{2}t_{c}^{4/3\gamma}t^{s_{3}}, \label{P55-4}%
\end{equation}%
\begin{equation}
\mu=C_{\mu}^{\left(  0\right)  }+\alpha_{\mu1}^{\left(  0\right)  }k^{2}%
t_{c}^{4/3\gamma}t^{s_{3}}, \label{P55-5}%
\end{equation}%
\begin{equation}
A=q_{1}C_{\lambda}^{\left(  1\right)  }t^{s_{1}-2}+q_{2}C_{\lambda}^{\left(
2\right)  }t^{s_{2}-2}+\alpha_{A1}^{\left(  0\right)  }k^{2}t_{c}^{4/3\gamma
}t^{s_{3}-2}. \label{P55-6}%
\end{equation}
where
\begin{equation}
s_{3}=\frac{2\left(  3\gamma-2\right)  }{3\gamma} \label{P55-7}%
\end{equation}

The exact coincidence of the forms of equations (\ref{P34})-(\ref{P36}) means
that in the main approximation the other pairs of amplitudes $\sigma_{\left(
J\right)  },B_{\left(  J\right)  }$ and $\omega_{\left(  J\right)
},D_{\left(  J\right)  }$ are described by the same formulas (\ref{P47}%
)-(\ref{P48}) and (\ref{P51})-(\ref{P52}) with only difference that instead of
$C_{\lambda}^{\left(  0\right)  },C_{\lambda}^{\left(  1\right)  },C_{\lambda
}^{\left(  2\right)  }$ one should take the new arbitrary constants
$C_{\sigma_{\left(  J\right)  }}^{\left(  0\right)  },C_{\sigma_{\left(
J\right)  }}^{\left(  1\right)  },C_{\sigma_{\left(  J\right)  }}^{\left(
2\right)  }$ and $C_{\omega_{\left(  J\right)  }}^{\left(  0\right)
},C_{\omega_{\left(  J\right)  }}^{\left(  1\right)  },C_{\omega_{\left(
J\right)  }}^{\left(  2\right)  }$ respectively. After that one can calculate
corrections to this main approximation taking into account the influence of
the terms in equations (\ref{P23})-(\ref{P25-A}) containing the factor
$k^{2}R^{-2}.$ These calculations are analogous to those we made for the
amplitudes $\lambda,\mu,t^{2}A$ and the final results are:
\begin{equation}
\sigma_{\left(  J\right)  }=C_{\sigma_{\left(  J\right)  }}^{\left(  0\right)
}+C_{\sigma_{\left(  J\right)  }}^{\left(  1\right)  }t^{s_{1}}+C_{\sigma
_{\left(  J\right)  }}^{\left(  2\right)  }t^{s_{2}},\text{ \ } \label{P56}%
\end{equation}%
\begin{equation}
\omega_{\left(  J\right)  }=C_{\omega_{\left(  J\right)  }}^{\left(  0\right)
}+C_{\omega_{\left(  J\right)  }}^{\left(  1\right)  }t^{s_{1}}+C_{\omega
_{\left(  J\right)  }}^{\left(  2\right)  }t^{s_{2}}+\alpha_{\omega_{\left(
J\right)  }1}^{\left(  0\right)  }k^{2}t_{c}^{4/3\gamma}t^{s_{3}},\text{ }
\label{P57}%
\end{equation}%
\begin{equation}
B_{\left(  J\right)  }=q_{1}C_{\sigma_{\left(  J\right)  }}^{\left(  1\right)
}t^{s_{1}-2}+q_{2}C_{\sigma_{\left(  J\right)  }}^{\left(  2\right)  }%
t^{s_{2}-2}, \label{P57-0}%
\end{equation}%
\begin{equation}
D_{\left(  J\right)  }=q_{1}C_{\omega_{\left(  J\right)  }}^{\left(  1\right)
}t^{s_{1}-2}+q_{2}C_{\omega_{\left(  J\right)  }}^{\left(  2\right)  }%
t^{s_{2}-2}+\alpha_{D_{\left(  J\right)  }1}^{\left(  0\right)  }k^{2}%
t_{c}^{4/3\gamma}t^{s_{3}-2}, \label{P57-1}%
\end{equation}
where the coefficients $\alpha_{\omega_{\left(  J\right)  }1}^{\left(
0\right)  }$ and $\alpha_{D_{\left(  J\right)  }1}^{\left(  0\right)  }$ are:
\begin{equation}
\alpha_{\omega_{\left(  J\right)  }1}^{\left(  0\right)  }=-\frac{9\gamma
^{2}\left(  3\gamma\beta-4\right)  }{2\left(  3\gamma-2\right)  \left[
24f+\left(  3\gamma+2\right)  \left(  3\gamma\beta-4\right)  \right]
}C_{\omega_{\left(  J\right)  }}^{\left(  0\right)  }\text{ }, \label{P57-2}%
\end{equation}%
\begin{equation}
\alpha_{D_{\left(  J\right)  }1}^{\left(  0\right)  }=-\frac{12f}{24f+\left(
3\gamma+2\right)  \left(  3\gamma\beta-4\right)  }C_{\omega_{\left(  J\right)
}}^{\left(  0\right)  }\text{ }. \label{P57-3}%
\end{equation}
Due to specific structure of equations (\ref{P23}) and (\ref{P24}) the
solutions for perturbations $\sigma_{\left(  J\right)  },$ $t^{2}B_{\left(
J\right)  }$ do not contain corrections of the order $t^{s_{3}}$.

The two arbitrary constants $C_{\sigma_{\left(  J\right)  }}^{\left(
0\right)  }$ can be removed by the coordinate transformations which still
remain in the synchronous system (in addition to those by which we already
eliminated constant $C_{\mu}^{\left(  -1\right)  }$ and can eliminate constant
$C_{\mu}^{\left(  0\right)  }$ in function $\mu$). Consequently the total
number of the arbitrary physical constants in the Fourier coefficients (which
generate the arbitrary 3-dimensional physical function in the real $x$-space)
of the solution is 13, these are $C_{\lambda}^{\left(  0\right)  },C_{\lambda
}^{\left(  1\right)  },C_{\lambda}^{\left(  2\right)  },$ $C_{\sigma_{\left(
J\right)  }}^{\left(  1\right)  },C_{\sigma_{\left(  J\right)  }}^{\left(
2\right)  },C_{\omega_{\left(  J\right)  }}^{\left(  0\right)  }%
,C_{\omega_{\left(  J\right)  }}^{\left(  1\right)  },C_{\omega_{\left(
J\right)  }}^{\left(  2\right)  }$. This is exactly the number of arbitrary
independent physical degrees of freedom of the system under consideration,
that is 4 for the gravitational field, 1 for the energy density, 3 for the
velocity and 5 for the shear stresses (five because the six components
$S_{\alpha\beta}$ follows from the six differential equations of the first
order in time with one additional condition $\delta^{\alpha\beta}%
S_{\alpha\beta}=0$). Then the solution we constructed is generic.

The asymptotic solutions for the Fourier coefficients for perturbations of the
velocity and energy density follow from (\ref{P26})-(\ref{P27}):%
\begin{align}
\tilde{V}_{\alpha}  &  =\frac{3ik\gamma}{8}\left(  \frac{2k_{\alpha}}%
{3k}C_{\lambda}^{\left(  1\right)  }-\sum_{J=2}^{3}l_{\alpha}^{\left(
J\right)  }C_{\sigma_{\left(  J\right)  }}^{\left(  1\right)  }\right)
s_{1}t^{s_{1}+1}+\label{P58}\\
&  +\frac{3ik\gamma}{8}\left(  \frac{2k_{\alpha}}{3k}C_{\lambda}^{\left(
2\right)  }-\sum_{J=2}^{3}l_{\alpha}^{\left(  J\right)  }C_{\sigma_{\left(
J\right)  }}^{\left(  2\right)  }\right)  s_{2}t^{s_{2}+1}+\nonumber\\
&  +\frac{ik_{\alpha}\left(  3\gamma-2\right)  }{6}\left(  \alpha_{\lambda
1}^{\left(  0\right)  }+\alpha_{\mu1}^{\left(  0\right)  }\right)  k^{2}%
t_{c}^{4/3\gamma}t^{s_{3}+1},\nonumber
\end{align}%
\begin{equation}
\tilde{E}=\frac{\gamma}{9\gamma-4}\left(  C_{\lambda}^{\left(  0\right)
}+C_{\mu}^{\left(  0\right)  }\right)  k^{2}t_{c}^{4/3\gamma}t^{s_{3}-2}.
\label{P59}%
\end{equation}

It is evident that the asymptotic behaviour of all perturbations satisfy the
basic requirement to be small in relative sense. This condition means that
variations (\ref{P11}) must be small with respect to the corresponding
background values, that is the quantities $\frac{\delta g_{\alpha\beta}}%
{R^{2}},\frac{\delta\varepsilon}{\varepsilon},\delta u_{\alpha}$ and
$\frac{\delta S_{\alpha\beta}}{\varepsilon R^{2}}$ in the limit $t\rightarrow
0$ should be much less than unity (the necessity to be small for the last
ratio follows from the condition $\delta S_{\alpha\beta}\ll T_{\alpha\beta
}^{\left(  0\right)  }=pg_{\alpha\beta}^{\left(  0\right)  }\sim\varepsilon
R^{2}$). In terms of the Fourier amplitudes these requirements are $\tilde
{H}_{\alpha\beta}\ll1,$ $t^{2}\tilde{E}\ll1,$ $\tilde{V}_{\alpha}\ll1,$
$t^{2}\tilde{K}_{\alpha\beta}\ll1$ and all of them are satisfied since all
time-dependent terms in the left hand sides of these inequalities are going to
die away as $t\rightarrow0$ and the six arbitrary constants
\begin{equation}
\tilde{H}_{\alpha\beta}^{\left(  0\right)  }=C_{\lambda}^{\left(  0\right)
}P_{\alpha\beta}\text{ }+C_{\mu}^{\left(  0\right)  }Q_{\alpha\beta}+%
{\displaystyle\sum\limits_{J=2}^{3}}
\left[  C_{\sigma_{\left(  J\right)  }}^{\left(  0\right)  }L_{\alpha\beta
}^{\left(  J\right)  }+C_{\omega_{\left(  J\right)  }}^{\left(  0\right)
}G_{\alpha\beta}^{\left(  J\right)  }\right]  \label{P60}%
\end{equation}
in the metric perturbations $\tilde{H}_{\alpha\beta}$ we are free to take to
be infinitesimally small. The interpretation of these constants is well known:
their appearance simply indicates that the isotropic part of the perturbed
metric $g_{\alpha\beta}$ in the $x$-space instead of the seed value
$R^{2}\delta_{\alpha\beta}$ acquires the more general form $R^{2}%
a_{\alpha\beta}\left(  x^{1},x^{2},x^{3}\right)  $ where $a_{\alpha\beta}$ in
perturbative solution should be closed to $\delta_{\alpha\beta}$ but in the
non-perturbative context (see below) becomes an arbitrary symmetric
3-dimensional tensor.

All this means that in the real $x$-space a generic non perturbative solution
exists with the following asymptotics for the metric near singularity:%
\begin{equation}
g_{\alpha\beta}=R^{2}\left(  a_{\alpha\beta}+t^{s_{1}}b_{\alpha\beta}^{\left(
1\right)  }+t^{s_{2}}b_{\alpha\beta}^{\left(  2\right)  }+t^{s_{3}}%
b_{\alpha\beta}^{\left(  3\right)  }+...\right)  \label{P61}%
\end{equation}
where $R=\left(  t/t_{c}\right)  ^{2/3\gamma}$ and exponents $s_{1},s_{2}$ and
$s_{3}$ are defined by the relation (\ref{P48}) and (\ref{P55-7}). The
additional terms denoted by the triple dots are small corrections which
contain the terms of the orders $t^{2s_{3}},t^{s_{1}+s_{3}},$ $t^{s_{2}+s_{3}%
}$ as well as all their powers and cross products. The main addend
$a_{\alpha\beta}$ represents six arbitrary 3-dimensional functions (in the
linearized version they are generated by the arbitrary constants $C_{\lambda
}^{\left(  0\right)  },C_{\mu}^{\left(  0\right)  },C_{\sigma_{\left(
J\right)  }}^{\left(  0\right)  },C_{\omega_{\left(  J\right)  }}^{\left(
0\right)  }$ in the Fourier coefficients). Each tensor $b_{\alpha\beta
}^{\left(  1\right)  }$ and $b_{\alpha\beta}^{\left(  2\right)  }$ consists of
the six 3-dimensional functions subjected to the restrictions $a^{\alpha\beta
}b_{\alpha\beta}^{\left(  1\right)  }=0$ and $a^{\alpha\beta}b_{\alpha\beta
}^{\left(  2\right)  }=0$ (here $a^{\alpha\beta}$ is inverse to $a_{\alpha
\beta}$), consequently $b_{\alpha\beta}^{\left(  1\right)  }$ and
$b_{\alpha\beta}^{\left(  2\right)  }$ contain another ten arbitrary
3-dimensional functions (in the linearized version they are generated by the
ten arbitrary constants $C_{\lambda}^{\left(  1\right)  },C_{\lambda}^{\left(
2\right)  },C_{\sigma_{\left(  J\right)  }}^{\left(  1\right)  }%
,C_{\sigma_{\left(  J\right)  }}^{\left(  2\right)  },C_{\omega_{\left(
J\right)  }}^{\left(  1\right)  },C_{\omega_{\left(  J\right)  }}^{\left(
2\right)  }$ in the Fourier coefficients). In case of complex conjugated
$s_{1}$ and $s_{2}$ the components $b_{\alpha\beta}^{\left(  1\right)  }$ and
$b_{\alpha\beta}^{\left(  2\right)  }$ are complex but in the way to provide
reality of the metric tensor. The last term $b_{\alpha\beta}^{\left(
3\right)  }$ and all corrections denoted by the triple dots in the expansion
(\ref{P61}) are expressible in terms of the $a_{\alpha\beta},b_{\alpha\beta
}^{\left(  1\right)  },b_{\alpha\beta}^{\left(  2\right)  }$ and their
derivatives then they do not contain any new arbitrariness. The shear
stresses, velocity and energy density follows from the exact Einstein
equations in terms of the metric tensor (\ref{P61}) and its derivatives and
all these quantities also do not contain any new arbitrary parameters. In
result the solution contains 16 arbitrary 3-dimensional functions the three of
which represent the gauge freedom due to the possibility of the arbitrary
3-dimensional coordinate transformations. Then the physical freedom in the
solution corresponds to 13 arbitrary functions as it should be.

This result is the generalization of the so-called quasi-isotropic solution
constructed in \cite{LK2} (see also \cite{LK}) for the perfect liquid.
However, in case of perfect liquid the isotropic singularity is unstable and
asymptotics found in \cite{LK2} corresponds to the narrow class of particular
solutions containing only 3 arbitrary physical 3-dimensional parameters.

\section{Concluding remarks}

1. The results presented show that the viscoelastic material with shear
viscosity coefficient $\eta\sim\sqrt{\varepsilon}$ can stabilize the Friedmann
cosmological singularity and the corresponding \textit{generic solution of the
Einstein equations for the viscous fluid possessing the isotropic Big Bang (or
Big Crunch) exists}. Depending on the free parameters $f,\beta,\gamma$ of the
theory such solution can be either of smooth power law asymptotics near
singularity (when both power exponents $s_{1}$ and $s_{2}$ are real and
positive) or it can have the character of damping (in the limit $t\rightarrow
0$) oscillations (when $s_{1}$ and $s_{2}$ have the positive real part and an
imaginary part). The last possibility reveals itself as a weak trace of the
chaotic oscillatory regime which is characteristic for the most general
asymptotics near the cosmological singularity and which can not be described
in closed analytical form (for the short simplified review on the oscillatory
regime see \cite{Bel1}). The present case show that the shear viscosity can
smooth such chaotic behaviour up to the quiet oscillations which have simple
asymptotic expressions in terms of the elementary functions of the type
$t^{\operatorname{Re}s}\sin\left[  \left(  \operatorname{Im}s\right)  \ln
t\right]  $ and $t^{\operatorname{Re}s}\cos\left[  \left(  \operatorname{Im}%
s\right)  \ln t\right]  .$

2. In the generic isotropic Big Bang described here some part of perturbations
are presented already at the initial singularity $t=0$ which are the three
physical components of the arbitrary 3-dimensional tensor $a_{\alpha\beta
}(x^{1},x^{2},x^{3})$ in formula (\ref{P61}). Another ten arbitrary physical
degrees of freedom are contained in the components of two tensors
$b_{\alpha\beta}^{\left(  1\right)  }$ and $b_{\alpha\beta}^{\left(  2\right)
}$ in this formula and they come to the action in the process of expansion.
This picture has no that shortage of the classical Lifshitz approach when one
is forced to introduce some unexplainable segment between singularity $t=0$
and initial time $t=t_{0}$ when perturbations arise in such a way that inside
this segment it is necessary to postulate without reasons the validity of the
exact Friedmann solution free of any perturbations.

3. It might happen that due to the universal growing of all perturbations (in
the course of expansion) already before that critical time when equations of
state will be changed and will switched off the action of viscosity the
perturbation amplitudes will reach the level sufficient for the further
development of the observed structure of our Universe. If not we always have
that means of escape as inflation phase which can be inserted in the evolution
after the Big Bang. Here we are touching another problem. It is known
\cite{Vil,Bor} that no inflation (including "eternal" one) can appear without
preceding cosmological singularity. Moreover, namely the period of expansion
from singularity to inflationary stage is responsible for the generation of
the necessary initial conditions for the such inflationary phase. How to match
the singular and inflationary stages and to find the initial conditions for
inflation call for another good piece of work.

4. In our analysis the case of stiff matter ($\gamma=2$) have been excluded.
This peculiar possibility should be investigated separately. It is known that
for the perfect liquid with stiff matter equation of state a generic solution
with isotropic singularity is impossible (see \cite{Bel1} and references
therein). The asymptotic of the general solution for this case have
essentially anisotropic structure although of the smooth (non-oscillatory)
power law character. It might be that viscosity will be able to isotropize
such evolution, however, it is not yet clear how the viscous stiff matter
should be treated mathematically. The simple way to take $\gamma=2$ in our
previous study does not works. \ 

5. Another interesting question is how an evolution directed outwards of a
thermally equilibrated state to a non-equilibrium one can be reconciled with
the second law of thermodynamics. Indeed, it seems that in accordance with
this law no deviation can happen from the background Friedmann expansion since
in course of a such deviation entropy must increase but in equilibrium it
already has the maximal possible value. The explanation should come from the
fact of the presence of the superstrong gravitational field. This field is an
external agent with respect to the matter itself, consequently, the matter in
the Friedmann Universe cannot be consider as closed system. It might happen
that Penrose \cite{Pen} is right and the gravitational field possess an
intrinsic entropy then this entropy being added to the entropy of matter will
bring the situation to the normal one. To clarify the question let's calculate
the matter entropy production near singularity in the solution described in
the previous sections. For the energy-momentum tensor (\ref{P1})-(\ref{P2})
equation $T_{i;k}^{k}u^{i}=0$ can be written as%
\begin{equation}
\left(  \sigma u^{k}\right)  _{;k}=-T^{-1}S^{mn}u_{m;n}\text{ }, \label{P62}%
\end{equation}
where $\sigma$ and $T$ is the entropy density and temperature of a (perturbed)
fluid. Here we used the fact that in our model chemical potential vanish (that
is $\gamma\varepsilon=T\sigma$) and that principal assumption of the
Israel-Stewart theory that the Gibbs relation (in our case $d\varepsilon
=Td\sigma$) is universal in the sense that it is valid for the arbitrary
displacements of the thermodynamical parameters, that is not only between
neighbouring equilibrium states. Equation (\ref{P4}) for stresses being
multiplied by $S^{ik}$ gives:%
\begin{equation}
S^{mn}u_{m;n}=-\frac{\tau}{2\eta}\left[  \frac{1}{2}\left(  S^{mn}%
S_{mn}\right)  _{;k}u^{k}+\frac{1}{\tau}S^{mn}S_{mn}\right]  . \label{P63}%
\end{equation}
Substituting this into the previous formula we obtain:%
\begin{gather}
\left(  \sigma u^{k}-\frac{\tau}{4\eta T}S^{mn}S_{mn}u^{k}\right)
_{;k}=\label{P64}\\
=\frac{1}{2\eta T}S^{mn}S_{mn}-\left(  \frac{\tau}{4\eta T}u^{k}\right)
_{;k}S^{mn}S_{mn}\nonumber
\end{gather}
The 4-vector in the brackets in the left hand side of the last equation
represents the generalization of the Landau-Lifshitz entropy flux for the case
when relaxation time $\tau$ of the shear stresses is not zero. This expression
for the entropy flux is the same that have been proposed by the Israel-Stewart
theory \cite{Isr1,Isr2}.

If the background solution is an real equilibrium state in the literal sense
then the action of the operator $u^{k}\partial_{k}$ on the
background\textit{\ }values of quantities $\tau,\eta,T$ gives zero and also
$u_{;k}^{k}=0$ for the background\textit{\ }values of the 4-velocity. Then the
factor $\left(  \tau u^{k}/4\eta T\right)  _{;k}$ in front of $S^{mn}S_{mn} $
in the last term of the equation (\ref{P64}) disappears in the first
approximation. Then this last term belongs to the third approximation since
also $S^{mn}$ vanish for the background solution. Consequently up to the
second order in the deviation from the equilibrium the equation (\ref{P64})
provides correct result, that is for any future directed evolution the entropy
increases because the quantity $S^{mn}S_{mn}$ is always positive due to the
properties (\ref{P2}) of the stresses.

However, the Friedmann background is not an equilibrium state in the
aforementioned literal sense. This solution describes the quasi-stationary
evolution in which the Universe passes the continuous sequence of equilibrium
states with different equilibrium parameters but with one and the same
conserved entropy. Due to this evolution the background value of the factor
$\left(  \tau u^{k}/4\eta T\right)  _{;k}$ in equation (\ref{P64}) is not
zero, moreover, it is not small with respect to the factor $1/2\eta T$ in the
first term in the right hand side of the equation (\ref{P64}). It is easy to
get $\left(  \tau u^{k}/4\eta T\right)  _{;k}$ from formulas (\ref{P9}%
)-(\ref{P10}) and (\ref{P45}) using expression $T=\gamma\varepsilon_{c}\left(
\varepsilon/\varepsilon_{c}\right)  ^{\left(  \gamma-1\right)  /\gamma}$ for
the background temperature ($\varepsilon_{c}$ is an arbitrary constant). In
result the entropy production equation (\ref{P64}) for our model take the form%
\begin{equation}
\left(  \sigma u^{k}-\frac{\tau}{4\eta T}S^{mn}S_{mn}u^{k}\right)  _{;k}%
=\frac{\beta-2}{2\beta\eta T}S^{mn}S_{mn}\text{ }, \label{P65}%
\end{equation}
and one can see that constant $\beta-2$ is negative. Indeed, the first
inequality in both sets of stability conditions (\ref{P49}) and (\ref{P50}) is
$\beta<\left(  3\gamma-2\right)  /\gamma$ but for any value of parameter
$\gamma$ from the interval $1\leqslant\gamma<2$ the quantity $\left(
3\gamma-2\right)  /\gamma$ is less than $2$.

It might be thought that the negativity of the right hand side of equation
(\ref{P65}) means that the second law of thermodynamics precludes the physical
realization of the generic isotropic Big Bang. However, it can happen that
such conclusion again would be too hasty because, as we already said, the
entropy of gravitational field might normalize the situation. As of now no
concrete calculation can be made inasmuch no theory of the gravitational
entropy exists. Nevertheless in the model under investigation it looks
plausible that gravitational entropy, being proportional to some invariants of
the Weyl tensor \cite{Pen}, indeed would be able to change the state of
affairs because for the background Friedmann solution this tensor is
identically zero and it will start to increase in the course of expansion.
Then increasing of the gravitational entropy would compensate the decreasing
of the matter entropy. For those who believe that the Universe began by an
isotropic expansion the negativity of the right hand side of the equation
(\ref{P65}) stands as a hint that gravitational entropy indeed exists.

By the way it is worth to remark that practically in all publications
(including \cite{Isr1,Isr2}) dedicated to the extended thermodynamics in the
framework of the General Relativity the condition $\sigma_{;k}^{k}\geqslant0$
for the entropy flux of the matter\textit{\ }is accepted from the beginning as
one's due. Moreover, namely from this condition follows the structure of the
additional dissipative terms in the energy-momentum tensor and particle flux.
Such strategy is undoubtedly correct not only for the "everyday life" but also
for the majority of the astrophysical problems where the gravitational fields
are relatively weak. However the cases with extremely strong gravity as in
vicinity to the cosmological singularity need more precise definition of what
we should understand under the total entropy of the system.

\section{Acknowledgements}

It is a pleasure to thank G.Bisnovatiy-Kogan for the useful critics and
stimulating discussions and E.Vladimirova for the help which accelerates the
creation of the final version of the manuscript.

\end{document}